# Sub-Bragg Phenomena in Multilayered Vibroacoustic Phononic Metamaterials

Majdi O. Gzal[a], Lawrence A. Bergman[b], Kathryn H. Matlack[a], Alexander F. Vakakis[a]

[a] Department of Mechanical Science and Engineering, University of Illinois, Urbana, USA

[b] Department of Aerospace Engineering, University of Illinois, Urbana, USA

## ABSTRACT

Beyond classical Bragg diffraction, we report on new sub-Bragg phenomena to achieve enhanced sound wave tunability in the low-frequency range in a multilayered acoustic metamaterial system. Remarkably, we reveal and study the formation of genuinely sub-wavelength Bragg-like band-splitting induced bandgaps, generalizing the "band-folding induced bandgaps" in the literature. Additionally, we propose a methodology to widen sub-wavelength local resonance bandgaps by simultaneously hosting two local resonances within the same bandgap. These sub-Bragg bandgaps are realized in an axisymmetric vibroacoustic phononic metamaterial consisting of repetitive multilayered unit cells, each composed of two layers of membrane-cavity resonators. The resulting coupled sound-structure interaction system is analytically solved in closed form. Furthermore, the studied multilayered vibroacoustic metamaterial exhibits sub-wavelength acoustical transparency, akin to "electromagnetically induced transparency", and an acoustic analogue of low-frequency "plasma oscillations", resulting in a plasma bandgap where wave propagation is entirely prohibited below the plasma frequency. These sub-wavelength phenomena, achieved predictively and well below the ubiquitous first-order Bragg diffraction range, provide broadband attenuation and superior wave manipulation capabilities for low-frequency sound. This highlights the potential of utilizing sound-structure interaction across diverse physical applications. Moreover, our findings can be extrapolated to wave propagation in broader classes of physical systems where sub-wavelength phenomena are crucial, including phononic and photonic crystals of more general configurations.

# I. INTRODUCTION

First discovered in the context of solid-state physics during the quantum revolution [1-6], *Bloch states* form the foundation for investigating how waves travel through structures with infinite periodically repeating patterns. Such infinite periodic media uniquely manipulate waves through their band structure, possessing pass bands and stop bands (or bandgaps), where waves either propagate (typically exhibiting strong dispersion) or exponentially decay (due to destructive interference), respectively. Analogous to *electronic crystals*, the band structure concept has been extended to electromagnetic waves, known as *photonic crystals* [7-10], as well as to acoustic and elastic waves, termed *phononic* (or *sonic*) *crystals* [11-22].

*Crystals* composed of an infinite number or repeating unit cells, whether *electronic*, *photonic*, or *phononic*, naturally support bandgaps due to their periodicity, achieved through *Bragg scattering*, which results from the destructive interference of *Bloch states* at unit-cell interfaces [23]. *Bragg diffraction* has led to significant advancements in these fields, enabling precise wave manipulation and tailoring, and resulting in improved electronic conduction [5,6], advanced optical communication [8,10], and superior sound and vibration control [14,24-26]. However, classical *Bragg diffraction* requires wavelengths comparable to the unit-cell size, restricting its application in *phononic crystals* at the ultrasonic regime [24-26] or necessitating bulky samples at the audible regime [14]. This limitation arises from the relatively large wavelengths of acoustic waves in the audible range, ranging from centimeters to meters.

To address the limitations of *Bragg diffraction* in *phononic crystals*, Liu et al. [27] introduced the concept of *local resonance bandgaps*. These bandgaps result from interactions between the wave field and local resonances within the crystal, where unlike *Bragg bandgaps*, are sub-wavelength and occur around resonant frequencies regardless of the underlying structural periodicity. This enables the formation of so-called *local resonance bandgaps* at much lower frequencies compared to the Bragg bandgaps, although their widths are dependent on the structural periodicity, and tend to be narrower when the unit-cell's natural frequency is significantly below the first *Bragg* frequency.

Several studies [28-38] have explored the synergy between Bragg scattering and local resonance bandgaps, revealing enhanced bandgap widths for effective wave attenuation. However, practical applications in low-frequency acoustic wave manipulation face distinct challenges, including the non-trivial overlap between local resonances and Bragg frequencies, necessitating bulky periodic structures. Additionally, implementing suitable materials for low-frequency manipulation, e.g., composed of soft and heavy coupled oscillators, poses other practical challenges. These complexities highlight significant engineering and scientific hurdles in employing phononic waveguides for tailoring acoustic waves at low frequencies, a task that is crucial for effectively mitigating low-frequency sound—a persistent environmental issue due to its penetrating power and impact on comfort and health [39-41].

As a promising solution to the challenge of low-frequency sound attenuation, significant efforts have been dedicated to developing various types of acoustic metamaterials [42-44], with particular emphasis on thin, lightweight membrane-type acoustic phononic systems. These metamaterials exhibit remarkable sound insulation capabilities due to their negative effective mass densities [45-50]. The effectiveness of membrane-type acoustic metamaterials in controlling and tailoring sound waves at low frequencies has been predominantly demonstrated through experiments and finite element analysis. Recently, the current authors proposed an approach [51] to develop exact single-term analytical solutions for coupled sound-structure interaction phononic systems and applied it in [52] to explore the limitations and capabilities of vibroacoustic metamaterials composed of monolayered membrane-cavity unit cells. Analytical results indicated that below the first-order *Bragg diffraction* limit, attenuation capabilities are constrained, particularly for the two leading pass bands within the sub-wavelength domain.

In the present study, we report unique *sub-Bragg* attenuation performance realized by extending the monolayered vibroacoustic metamaterial in [52] to more general multilayered axisymmetric configurations. We solve the coupled multilayered sound-membrane problem in closed-form, enabling an in-depth predictive study of the band structure of these systems. This examination reveals novel sub-Bragg mechanisms occurring well below the first-order Bragg diffraction, including the formation of genuinely *sub-wavelength Bragg-like band-splitting induced bandgaps* and the widening of *sub-wavelength local resonance bandgaps*. Furthermore, our findings reveal acoustic analogues of "plasma oscillations" as well as "electromagnetically induced transparency", occurring within the sub-wavelength range. These findings highlight the potential of multilayered configurations in achieving superior sound attenuation and manipulation at low frequencies. Additionally, in the realm of topological phase transition, the "zone-folding" strategy has been utilized [53-55] to generate "band-folding induced bandgaps" or "zone-folding-induced bandgaps" [56-58]. Such bandgaps can emerge in the sub-wavelength range and their formation occurs in the vicinities of degenerate points caused by the band folding effect. In this study, we show that the proposed *band-splitting induced bandgap* concept provides a general class of sub-wavelength bandgaps, including "band-folding induced bandgaps" as a special case. Unlike local resonance bandgaps, this broad family of sub-wavelength band-splitting bandgaps exhibits Bragg-like attenuation capabilities and supports the formation of topological interface states, as recently demonstrated by the authors in [59]. Therefore, our findings open new avenues for exploring the nontrivial topological properties of band-splitting bandgaps.

## II. ANALYTICAL TREATMENT OF THE INFINITE PERIODIC VIBROACOUSTIC METAMATERIAL LATTICE USING BLOCH'S THEOREM

We consider an axisymmetric cylindrical duct embedded with an infinite number of multilayered unit cells composed of coupled vibroacoustic resonators. The proposed configuration forms a phononic vibroacoustic metamaterial with a fundamental multilayered unit cell periodically repeated in space to create a periodic lattice. A schematic representation of a periodic assembly of three adjacent unit-cells is illustrated in Figure 1(a). The multilayered unit cell consists of two layers of membrane-cavity resonators. Specifically, it comprises a sequence of Membrane A, Cavity 1, Membrane B, and Cavity 2. A representative unit-cell is shown in Figure 1(b).

In such an infinite periodic lattice, *Bloch's theorem* can be employed to analyze the propagation of sound waves between adjacent cells, based on the assumption that the change in wave amplitude between cells is independent of their specific locations within the lattice. Consequently, a thorough analysis of the reference unit-cell enables a complete characterization of the wave propagation characteristics in the infinite periodic assembly.

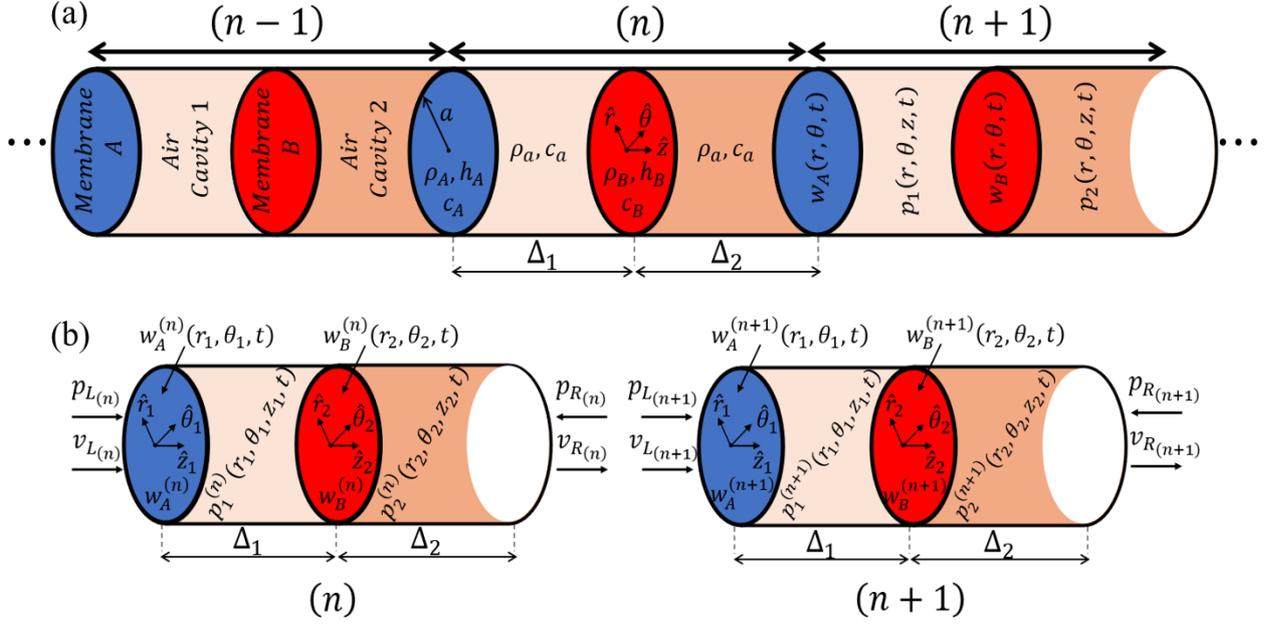

**Figure 1:** Schematic of the infinite vibroacoustic multilayered metamaterial: (a) Three adjacent unit-cells, and (b) the $n^{th}$ and $(n+1)^{th}$ unit-cells showing acoustic velocities and pressures at their left ($L$) and right ($R$) boundaries; Blue (red) circles represent Membrane A (B), while light (dark) orange indicates the acoustic air Cavity 1 (2), respectively.

### A. Analyzing a representative multilayered vibroacoustic unit-cell

Two local cylindrical coordinate systems, represented by $(\hat{r}_1, \hat{\theta}_1, \hat{z}_1)$ and $(\hat{r}_2, \hat{\theta}_2, \hat{z}_2)$, located at the center of Membranes A and B, respectively, are considered for analyzing the $n^{th}$ unit-cell (see Figure 2(b)). The duct radius, denoted by $a$, is the same for both the circular membranes and the cylindrical cavities. The depths of Cavity 1 and Cavity 2 are represented by $\Delta_1$ and $\Delta_2$, respectively. The density and the sound speed of the air inside the two cavities are denoted by $\rho_a$ and $c_a$, respectively. The membranes are assumed to be clamped along their edges, with each membrane possessing a uniform thickness, material density, speed of elastic wave propagation, and viscous damping; these properties are denoted as $h_A$, $\rho_A$, $c_A$, and $\lambda_A$ for Membrane A, and $h_B$, $\rho_B$, $c_B$, and $\lambda_B$ for Membrane B, respectively. The thicknesses of the membranes are assumed to be negligible in comparison to the depths of the cavities, i.e., $h_A, h_B \ll \Delta_1, \Delta_2$; consequently, the characteristic length of the unit-cell is simply equal to $\Delta_1 + \Delta_2$. In the following derivations, we will assume that the amplitudes of the transversal vibrations of the membranes are small enough for the linearized form of infinitesimal elasticity to apply.

The $n^{th}$ unit-cell is isolated from its neighboring cells, and its free body equilibrium is illustrated in Figure 1(b). The interaction between the cells is described using the acoustic pressure and velocity denoted by $p_{L_{(n)}}$ and $v_{L_{(n)}}$ at the left boundary ($L$), and $p_{R_{(n)}}$ and $v_{R_{(n)}}$ at the right boundary ($R$), respectively. We assume that the acoustic pressure and velocity at the left boundary of the $n^{th}$ unit-cell are known quantities, while the acoustic state (i.e., pressure and velocity) at the right boundary are unknown.

Denote by $p_1^{(n)}(r_1, \theta_1, z_1, t)$ and $p_2^{(n)}(r_2, \theta_2, z_2, t)$ the instantaneous acoustic pressure distributions within Cavities 1 and 2, respectively, inside the $n^{th}$ unit-cell . These pressure distributions are governed by the following homogeneous 3D wave equations represented in cylindrical coordinates [60]:

$$c_a^2 \left( \frac{\partial^2 p_1^{(n)}}{\partial r_1^2} + \frac{1}{r_1} \frac{\partial p_1^{(n)}}{\partial r_1} + \frac{1}{r_1^2} \frac{\partial^2 p_1^{(n)}}{\partial \theta_1^2} + \frac{\partial^2 p_1^{(n)}}{\partial z_1^2} \right) = \frac{\partial^2 p_1^{(n)}}{\partial t^2}$$
$$c_a^2 \left( \frac{\partial^2 p_2^{(n)}}{\partial r_2^2} + \frac{1}{r_2} \frac{\partial p_2^{(n)}}{\partial r_2} + \frac{1}{r_2^2} \frac{\partial^2 p_2^{(n)}}{\partial \theta_2^2} + \frac{\partial^2 p_2^{(n)}}{\partial z_2^2} \right) = \frac{\partial^2 p_2^{(n)}}{\partial t^2}$$
(1)

To implement the appropriate boundary conditions for the acoustic pressures inside the cavities, we utilize the linearized Euler transport equation, which establishes the following relationship between acoustic pressure ($p$) and velocity ($v$):

$$v = -\frac{1}{\rho_a} \int \frac{\partial p}{\partial z} dt \qquad (2)$$

The cylindrical duct is assumed to be acoustically rigid in the radial direction, meaning that the radial component of the flow velocity at the duct's circumference is required to be zero. Therefore, according to the linearized Euler transport equation, the acoustic pressures inside the cavities satisfy the following boundary conditions in the radial direction:

$$\left. \frac{\partial p_1^{(n)}}{\partial r_1} \right|_{r_1=a} = 0, \quad \left. \frac{\partial p_2^{(n)}}{\partial r_2} \right|_{r_2=a} = 0 \qquad (3)$$

In addition, the continuity of the normal acoustic velocity (along the $\hat{z}$ direction) at the cavity-membrane interface necessitates that:

$$\left. \frac{\partial p_1^{(n)}}{\partial z_1} \right|_{z_1=0} = -\rho_a \frac{\partial^2 w_A^{(n)}}{\partial t^2}, \quad \left. \frac{\partial p_1^{(n)}}{\partial z_1} \right|_{z_1=\Delta_1} = -\rho_a \frac{\partial^2 w_B^{(n)}}{\partial t^2},$$
$$-\frac{1}{\rho_a} \left[ \int \frac{\partial p_1^{(n)}}{\partial z_1} dt \right]_{z_1=0} = v_{L_n}$$
$$\left. \frac{\partial p_2^{(n)}}{\partial z_2} \right|_{z_2=0} = -\rho_a \frac{\partial^2 w_B^{(n)}}{\partial t^2}, \quad \left. p_2^{(n)} \right|_{z_2=\Delta_2} = -p_{R_n},$$
$$-\frac{1}{\rho_a} \left[ \int \frac{\partial p_2^{(n)}}{\partial z_2} dt \right]_{z_2=\Delta_2} = v_{R_n}$$
(4)

Here, $w_A^{(n)}(r_1, \theta_1, t)$ and $w_B^{(n)}(r_2, \theta_2, t)$ represent the instantaneous transverse displacements of Membrane A and Membrane B, respectively, inside the $n^{th}$ unit-cell. These forced damped transverse displacements of the membranes are governed by the following inhomogeneous 2D wave equations represented in polar coordinates,

$$c_A^2 \left( \frac{\partial^2 w_A^{(n)}}{\partial r_1^2} + \frac{1}{r_1} \frac{\partial w_A^{(n)}}{\partial r_1} + \frac{1}{r_1^2} \frac{\partial^2 w_A^{(n)}}{\partial \theta_1^2} \right) + \frac{1}{\rho_A h_A} \left( p_{L_n} - p_1^{(n)} \big|_{z_1=0} \right)$$
$$= \frac{\partial^2 w_A^{(n)}}{\partial t^2} + \frac{\lambda_A}{\rho_A h_A} \frac{\partial w_A^{(n)}}{\partial t}$$
(5)

$$c_B^2 \left( \frac{\partial^2 w_B^{(n)}}{\partial r_2^2} + \frac{1}{r_2} \frac{\partial w_B^{(n)}}{\partial r_2} + \frac{1}{r_2^2} \frac{\partial^2 w_B^{(n)}}{\partial \theta_2^2} \right) + \frac{1}{\rho_B h_B} \left( p_1^{(n)} \bigg|_{z_1=\Delta_1} - p_2^{(n)} \bigg|_{z_2=0} \right)$$

$$= \frac{\partial^2 w_B^{(n)}}{\partial t^2} + \frac{\lambda_B}{\rho_B h_B} \frac{\partial w_B^{(n)}}{\partial t}$$

with clamped boundary conditions:

$$w_A^{(n)} \bigg|_{r_1=a} = 0, \quad w_B^{(n)} \bigg|_{r_2=a} = 0 \tag{6}$$

In this study, we assume that the longitudinal modes of both Cavities 1 and 2, as well as the transversal modes of Membranes A and B, are axisymmetric, so the acoustic pressure distributions inside the cavities and the transverse displacement of the membranes are independent of $\theta$, i.e., satisfying $\frac{\partial (\cdot)}{\partial \theta} = 0$.

Focusing on the two Cavities, the steady-state axisymmetric acoustic pressures for fixed frequency $\Omega$ that satisfy the equation of motion (1) and the boundary conditions (3) are expressed as,

$$p_1^{(n)}(r_1, z_1, t) = \sum_{m=0}^{\infty} J_0(r_1 \beta_m) \left( P_{1c}^{(m)} \cos(z_1 k_m) + P_{1s}^{(m)} \sin(z_1 k_m) \right) e^{i\Omega t}$$

$$p_2^{(n)}(r_2, z_2, t) = \sum_{m=0}^{\infty} J_0(r_2 \beta_m) \left( P_{2c}^{(m)} \cos(z_2 k_m) + P_{2s}^{(m)} \sin(z_2 k_m) \right) e^{i\Omega t} \tag{7}$$

where $J_0$ is the zeroth order Bessel function of the first kind, $P_{jc}^{(m)}$ and $P_{js}^{(m)}$ are the Cosine and Sine modal amplitudes, respectively, of the $m^{th}$ mode for $p_j^{(n)}$, $j = 1,2$, $k_m$ is the $m^{th}$ mode wave number, and $\beta_m$ is the $m^{th}$ root of the first order Bessel function of the first kind, $J_1$, divided by the membrane radius $a$, i.e., $J_1(a\beta_m) = 0$. The relationship between $k_m$ and $\beta_m$ is given by:

$$k_m = \sqrt{\frac{\Omega^2}{c_a^2} - \beta_m^2} \tag{8}$$

It is evident that the modal wave number $k_m$ can be either real or imaginary. Real values indicate propagating (volume) modes, whereas imaginary values correspond to localized (surface) modes, which are spatially confined near the membrane surface. Therefore, the acoustic pressure time series (7) can be truncated to a finite number of harmonics, denoted as $N_m$, based on the *modal cut-on frequency*, denoted by $\Omega_m^{cut-on}$. This modal cut-on frequency determines the minimum threshold below which there is no propagation of the $m^{th}$ mode, and it is given by, $\Omega_m^{cutoff} = \beta_m c_a = \frac{\chi_m}{a} c_a$, where $\chi_m$ is the $m^{th}$ root of the first order Bessel function of the first kind, $J_1$. In this study, we will fix the sound speed of the air to its actual value at room temperature, i.e., $c_a = 343 \, [m/s]$ and the membrane (or duct) radius to $a = 5 \, [cm]$. Consequently, the cut-on frequency for the case of $m = 1$ is equal to $\Omega_1^{cutoff} = 4,183 \, [Hz]$. Therefore, in the frequency range of $0 < \Omega < \Omega_1^{cutoff} = 4,183 \, [Hz]$, the infinite series of acoustic pressure distributions given in (4) can be approximately truncated to a single harmonic with $m = 0$, i.e., there is a state of monoharmonic coupling, as follows,

$$p_1^{(n)}(z_1, t) = \left( P_{1c}^{(0)} \cos(z_1 k_0) + P_{1s}^{(0)} \sin(z_1 k_0) \right) e^{i\Omega t} \tag{9}$$

$$p_2^{(n)}(z_2, t) = \left(P_{2c}^{(0)}\cos(z_2 k_0) + P_{2s}^{(0)}\sin(z_2 k_0)\right) e^{i\Omega t}$$

where $k_0 = \Omega/c_a$. These pressure distributions are represented as uniform plane waves propagating in the $\hat{z}$ direction and couple the elastic membranes through a single harmonic. Accordingly, the acoustic pressure and velocity at the left and right boundaries of the $n^{th}$ unit-cell can be represented as plane waves,

$$\begin{aligned} p_{L_{(n)}} &= P_{L_n} e^{i\Omega t}, \; v_{L_{(n)}} = V_{L_n} e^{i\Omega t} \\ p_{R_{(n)}} &= P_{R_n} e^{i\Omega t}, \; v_{R_{(n)}} = V_{R_n} e^{i\Omega t} \end{aligned} \quad (10)$$

where $P_{L_n}$ and $V_{L_n}$ are treated as knowns, while $P_{R_n}$ and $V_{R_n}$ as unknowns.

By introducing the pressure distributions for the cavities (9) and the pressure at the unit-cell edges (10) into the equations of motion for the membranes (5) and applying the axisymmetric assumption, we obtain:

$$\begin{aligned} c_A^2 \left( \frac{\partial^2 w_A^{(n)}}{\partial r_1^2} + \frac{1}{r_1} \frac{\partial w_A^{(n)}}{\partial r_1} \right) + \frac{1}{\rho_A h_A} \left(P_{L_n} - P_{1c}^{(0)}\right) e^{i\Omega t} \\ = \frac{\partial^2 w_A^{(n)}}{\partial t^2} + \frac{\lambda_A}{\rho_A h_A} \frac{\partial w_A^{(n)}}{\partial t} \\ c_B^2 \left( \frac{\partial^2 w_B^{(n)}}{\partial r_2^2} + \frac{1}{r_2} \frac{\partial w_B^{(n)}}{\partial r_2} \right) \\ + \frac{1}{\rho_B h_B} \left(P_{1c}^{(0)}\cos(\Delta_1 k_0) + P_{1s}^{(0)}\sin(\Delta_1 k_0) - P_{2c}^{(0)}\right) e^{i\Omega t} \\ = \frac{\partial^2 w_B^{(n)}}{\partial t^2} + \frac{\lambda_B}{\rho_B h_B} \frac{\partial w_B^{(n)}}{\partial t} \end{aligned} \quad (11)$$

Instead of using the conventional method of representing the transverse displacement of the membranes through an infinite series of eigenfunctions that satisfy the clamped boundary conditions, this study utilizes a recent approach [51,52] which directly solves the transverse displacements of the membranes without assuming eigenfunction (modal) expansions. Specifically, the forced steady-state transverse deformations of the membranes are approached through classical space-time separation. This procedure is followed by solving nonhomogeneous ordinary differential equations in terms of superpositions of particular and homogeneous solutions. Finally, the regularity conditions, i.e., $\left|w_A^{(n)}\right|_{r_1=0}, \left|w_B^{(n)}\right|_{r_2=0} < \infty$, and the clamped boundary conditions, i.e., $w_A^{(n)}\big|_{r_1=a} = w_B^{(n)}\big|_{r_2=a} = 0$, are applied, resulting in the following exact solution for the transverse displacements of the forced damped membranes:

$$\begin{aligned} w_A^{(n)}(r_1, t) &= \frac{\left(P_{1c}^{(0)} - P_{L_n}\right)}{\eta_A^2 \rho_A h_A c_A^2} \left(1 - \frac{J_0(r_1 \eta_A)}{J_0(a \eta_A)}\right) e^{i\Omega t} \\ w_B^{(n)}(r_2, t) &= \frac{\left(P_{2c}^{(0)} - P_{1c}^{(0)}\cos(\Delta_1 k_0) - P_{1s}^{(0)}\sin(\Delta_1 k_0)\right)}{\eta_B^2 \rho_B h_B c_B^2} \left(1 - \frac{J_0(r_2 \eta_B)}{J_0(a \eta_B)}\right) e^{i\Omega t} \\ \eta_A &= \sqrt{\frac{\Omega^2}{c_A^2} - \frac{i\Omega \lambda_A}{\rho_A h_A c_A^2}}, \quad \eta_B = \sqrt{\frac{\Omega^2}{c_B^2} - \frac{i\Omega \lambda_B}{\rho_B h_B c_B^2}} \end{aligned} \quad (12)$$

These transverse displacement distributions of the membranes are expressed in terms of the amplitudes of the acoustic pressure inside the cavities, i.e., $P_{1c}^{(0)}$, $P_{1s}^{(0)}$, $P_{2c}^{(0)}$, and $P_{2s}^{(0)}$, which remain unknown at this stage. To fully characterize the vibroacoustic state of the $n^{th}$ unit-cell, one needs to solve for these four unknowns as well as for the acoustic pressure and velocity at the right boundary ($P_{R_n}$ and $V_{R_n}$ in equation (10)). This can be accomplished by substituting equations (9), (10) and (12) into the six interface boundary conditions given in equation (4). After rather lengthy, yet elementary, algebraic manipulations, the following expressions for the unknown amplitudes are obtained in terms of the acoustic pressure and velocity at the left boundary:

$$P_{1c}^{(0)} = P_{L_n} + \frac{i}{\Omega}\sigma_A V_{L_n}, \quad P_{1s}^{(0)} = -\frac{i}{\Omega}\sigma_a V_{L_n}$$

$$P_{2c}^{(0)} = \left[\cos(k_0\Delta_1) + \sin(k_0\Delta_1)\frac{\sigma_B}{\sigma_a}\right]P_{L_n}$$
$$+ \frac{i}{\Omega}\left[\cos(k_0\Delta_1)(\sigma_A + \sigma_B) + \sin(k_0\Delta_1)\left(\frac{\sigma_A\sigma_B}{\sigma_a} - \sigma_a\right)\right]V_{L_n}$$

$$P_{2s}^{(0)} = -\sin(k_0\Delta_1)P_{L_n} - \frac{i}{\Omega}[\sin(k_0\Delta_1)\sigma_A + \cos(k_0\Delta_1)\sigma_a]V_{L_n}$$

$$P_{R_n} = -\left[\cos(k_0(\Delta_1 + \Delta_2)) + \sin(k_0\Delta_1)\cos(k_0\Delta_2)\frac{\sigma_B}{\sigma_a}\right]P_{L_n}$$
$$+ \frac{i}{\Omega}\Big[\sin(k_0(\Delta_1 + \Delta_2))\sigma_a - \cos(k_0(\Delta_1 + \Delta_2))\sigma_A \quad (13)$$
$$- \cos(k_0\Delta_2)\sigma_B\left(\cos(k_0\Delta_1) + \sin(k_0\Delta_1)\frac{\sigma_A}{\sigma_a}\right)\Big]V_{L_n}$$

$$V_{R_n} = \frac{\Omega}{i\sigma_a}\Big[\sin(k_0(\Delta_1 + \Delta_2)) + \sin(k_0\Delta_1)\sin(k_0\Delta_2)\frac{\sigma_B}{\sigma_a}\Big]P_{L_n}$$
$$+ \Big[\cos(k_0(\Delta_1 + \Delta_2)) + \sin(k_0(\Delta_1 + \Delta_2))\frac{\sigma_A}{\sigma_a}$$
$$+ \sin(k_0\Delta_2)\frac{\sigma_B}{\sigma_a}\left(\cos(k_0\Delta_1) + \sin(k_0\Delta_1)\frac{\sigma_A}{\sigma_a}\right)\Big]V_{L_n}$$

where,

$$\sigma_A = \eta_A^2\rho_A h_A c_A^2 \frac{J_0(a\eta_A)}{J_2(a\eta_A)}, \quad \sigma_B = \eta_B^2\rho_B h_B c_B^2 \frac{J_0(a\eta_B)}{J_2(a\eta_B)}, \quad \sigma_a = \frac{\rho_a\Omega^2}{k_0} = \rho_a c_a \Omega$$

Finally, by substituting the pressure distributions amplitudes (13) into (9) and (12), the pressure distributions inside the cavities and the membrane transverse displacements become:

$$p_1^{(n)}(z_1, t) = \left(\left(P_{L_n} + \frac{i}{\Omega}\sigma_A V_{L_n}\right)\cos(z_1 k_0) - \frac{i}{\Omega}\sigma_a V_{L_n}\sin(z_1 k_0)\right)e^{i\Omega t}$$

$$p_2^{(n)}(z_2, t) = \Big(\cos(k_0\Delta_1)\Big(P_{L_n} + \tan(k_0\Delta_1)\frac{\sigma_A}{\sigma_a}P_{L_n}$$
$$+ \frac{i}{\Omega}\left(\sigma_A + \sigma_B + \tan(k_0\Delta_1)\left(\frac{\sigma_A\sigma_B}{\sigma_a} - \sigma_a\right)\right)V_{L_n}\Big)\cos(z_2 k_0) \quad (14)$$
$$+ \sin(k_0\Delta_1)\left(-P_{L_n} - \frac{i}{\Omega}(\sigma_A + \cot(k_0\Delta_1)\sigma_a)V_{L_n}\right)\sin(z_2 k_0)\Big)e^{i\Omega t}$$

$$w_A^{(n)}(r_1, t) = \frac{i}{\Omega}V_{L_n}\left(\frac{J_0(a\eta_A) - J_0(r_1\eta_A)}{J_2(a\eta_A)}\right)e^{i\Omega t}$$

$$w_B^{(n)}(r_2, t) = sin(k_0\Delta_1)\left(\frac{1}{\sigma_a}P_{Ln}\right.$$
$$\left. + \frac{i}{\Omega}\left(cot(k_0\Delta_1) + \frac{\sigma_A}{\sigma_a}\right)V_{Ln}\right)\left(\frac{J_0(a\eta_B) - J_0(r_2\eta_B)}{J_2(a\eta_B)}\right)e^{i\Omega t}$$

It is important to note that when eliminating the effect of Cavity 1 on Membrane A, i.e., by substituting $P_{1c}^{(0)} = 0$ in equation (12), the in-vacuo membrane natural frequencies are obtained by solving the equation $J_0(a\eta_A) = 0$. However, in the presence of Cavity 1, the natural frequencies of Membrane A become the roots of $J_2(a\eta_A)$, as shown in (14).

## B. Constructing the local transfer, admittance and impedance matrices

Let $\mathbf{y}_{L(n)}$ and $\mathbf{y}_{R(n)}$ represent the acoustic state vectors at the left and right boundaries, respectively, of the $n^{th}$ unit-cell and are defined as:

$$\mathbf{y}_{L(n)} = \begin{bmatrix} \frac{p_{L(n)}}{\rho_a c_a} \\ v_{L(n)} \end{bmatrix}, \quad \mathbf{y}_{R(n)} = \begin{bmatrix} \frac{p_{R(n)}}{\rho_a c_a} \\ v_{R(n)} \end{bmatrix} \quad (15)$$

The acoustic velocity compatibility and the pressure equilibrium at the junction between the $n^{th}$ and the $(n+1)^{th}$ unit-cells, as shown in Figure 1(b), suggest that:

$$\mathbf{y}_{L(n+1)} = \begin{bmatrix} \frac{p_{L(n+1)}}{\rho_a c_a} \\ v_{L(n+1)} \end{bmatrix} = \begin{bmatrix} -\frac{p_{R(n)}}{\rho_a c_a} \\ v_{R(n)} \end{bmatrix} \quad (16)$$

Now, we construct the *local transfer matrix*, denoted by $\mathbf{T}$, which represents the recursion relation between the acoustic state vectors at the left boundaries of the $n^{th}$ and $(n+1)^{th}$ unit-cells as follows:

$$\mathbf{y}_{L(n+1)} = \mathbf{T}\mathbf{y}_{L(n)} \quad (17)$$

where the components of the *local transfer matrix*, $\mathbf{T}$, which are frequency dependent, are explicitly given by:

$$\mathbf{T} = \begin{bmatrix} T_{11} & T_{12} \\ T_{21} & T_{22} \end{bmatrix}$$
$$T_{11} = cos(k_0(\Delta_1 + \Delta_2)) + sin(k_0\Delta_1)cos(k_0\Delta_2)\frac{\sigma_B}{\sigma_a}$$
$$T_{12} = i\left[-sin(k_0(\Delta_1 + \Delta_2)) + cos(k_0(\Delta_1 + \Delta_2))\frac{\sigma_A}{\sigma_a}\right.$$
$$\left. + cos(k_0\Delta_2)\frac{\sigma_B}{\sigma_a}\left(cos(k_0\Delta_1) + sin(k_0\Delta_1)\frac{\sigma_A}{\sigma_a}\right)\right] \quad (18)$$
$$T_{21} = -i\left[sin(k_0(\Delta_1 + \Delta_2)) + sin(k_0\Delta_1)sin(k_0\Delta_2)\frac{\sigma_B}{\sigma_a}\right]$$
$$T_{22} = cos(k_0(\Delta_1 + \Delta_2)) + sin(k_0(\Delta_1 + \Delta_2))\frac{\sigma_A}{\sigma_a}$$
$$+ sin(k_0\Delta_2)\frac{\sigma_B}{\sigma_a}\left(cos(k_0\Delta_1) + sin(k_0\Delta_1)\frac{\sigma_A}{\sigma_a}\right)$$

The components of the *local transfer matrix*, $\mathbf{T}$, are derived by substituting equation (13) into equation (16), and then incorporating the results into equation (17). Two adjacent unit cells are interconnected by the acoustic pressure and velocity within the cavities, represented by a single harmonic (see equation (9)). Consequently, the proposed lattice comprises mono-coupled unit cells, making the local transfer matrix, $\boldsymbol{T}$, a $2 \times 2$ matrix. In a multi-coupled system (for high frequency operation, $\Omega > \Omega_1^{cutoff} = \frac{\chi_1}{a} c_a = 4183\ [Hz]$), the transfer matrix components become submatrices. Normalizing the acoustic pressure in the state vector (15) by the factor $\rho_a c_a$ ensures that the elements of the local transfer matrix are nondimensional.

Rearranging the terms in equation (17) yields a relation between the acoustic velocities at the left and right boundaries of the $n^{th}$ unit cell to the acoustic pressures at the same locations. This relation is expressed using a *local admittance matrix*, $\boldsymbol{Y}$, as follows:

$$\begin{bmatrix} v_{L_{(n)}} \\ v_{R_{(n)}} \end{bmatrix} = \boldsymbol{Y} \begin{bmatrix} \dfrac{p_{L_{(n)}}}{\rho_a c_a} \\ \dfrac{p_{R_{(n)}}}{\rho_a c_a} \end{bmatrix}$$

where,

$$\boldsymbol{Y} = \begin{bmatrix} Y_{11} & Y_{12} \\ Y_{21} & Y_{22} \end{bmatrix} = \begin{bmatrix} -\dfrac{T_{11}}{T_{12}} & -\dfrac{1}{T_{12}} \\ -\dfrac{(T_{11}T_{22} - T_{12}T_{21})}{T_{12}} & -\dfrac{T_{22}}{T_{12}} \end{bmatrix}$$

(19)

Therefore, the inverse of the *local admittance matrix* is the *local impedance matrix*, denoted by $\boldsymbol{Z}$, and expressed as:

$$\begin{bmatrix} \dfrac{p_{L_{(n)}}}{\rho_a c_a} \\ \dfrac{p_{R_{(n)}}}{\rho_a c_a} \end{bmatrix} = \boldsymbol{Z} \begin{bmatrix} v_{L_{(n)}} \\ v_{R_{(n)}} \end{bmatrix}, \quad \boldsymbol{Z} = \begin{bmatrix} Z_{11} & Z_{12} \\ Z_{21} & Z_{22} \end{bmatrix} = \boldsymbol{Y}^{-1}$$

(20)

The components of the *local admittance* and *impedance matrices*, $\boldsymbol{Y}$ and $\boldsymbol{Z}$, in (19) and (20), respectively, are expressed in terms of the components of *the local transfer matrix*, $\boldsymbol{T}$. As the considered system is time-invariant, linear, and possesses scalar material properties, the *reciprocity principle* holds [61], necessitating $Y_{12} = Y_{21}$ (or $Z_{12} = Z_{21}$). Based on equation (19) the symmetry of $\boldsymbol{Y}$ implies that the *local transfer matrix*, $\boldsymbol{T}$, must satisfy $det(\boldsymbol{T}) = T_{11}T_{22} - T_{12}T_{21} = 1$, i.e., $\boldsymbol{T}$ is a unimodular matrix. It can be proved that the components of the *transfer matrix* $\boldsymbol{T}$ given in (18), indeed satisfy the unimodularity condition regardless of $\Omega$. The matrix $\boldsymbol{T}$ defines how acoustic state amplitudes evolve across unit-cell boundaries with frequency. Its unimodularity ensures non-singularity, allowing for its inverse, $\boldsymbol{T}^{-1}$, obtained by interchanging state vectors in equation (17) and given by $\boldsymbol{T}^{-1} = adj(\boldsymbol{T})$. $\boldsymbol{T}^{-1}$ serves as the transfer matrix in the opposite direction, sharing identical eigenvalues and eigenvectors with $\boldsymbol{T}$. Physically, the unimodularity of the transfer matrix, i.e., the fact that this is a reciprocal system, implies identical spectral behavior in both left-to-right and right-to-left propagation directions.

Another perspective to completely describe the acoustical behavior of the infinite periodic lattice is to use the *local scattering matrix*, which encodes the scattering coefficients, including reflection, transmission, and absorption coefficients. Interested readers are referred to [52], where a general expression for the *scattering matrix* in terms of the *local transfer matrix* components and the scattering coefficients is derived.

## C. Band structure analysis

To explore wave propagation within the infinite lattice, we apply *Floquet's theorem*, linking the state vectors of consecutive unit cells through the expression:

$$\boldsymbol{y}_{L(n+1)} = \Lambda \boldsymbol{y}_{L(n)} \qquad (21)$$

Here $\Lambda$ represents the matrix of *Floquet multipliers*. By incorporating the transfer matrix relation from (17) into (21), we derive the following linear eigenvalue problem:

$$\boldsymbol{T}\boldsymbol{y}_{L(n)} = \Lambda \boldsymbol{y}_{L(n)} \qquad (22)$$

The eigenvalues of the transfer matrix, denoted by $\Lambda_1$ and $\Lambda_2$, known as the Floquet multipliers, describe the evolution of state vector amplitudes between two consecutive unit cells at a given frequency. Due to the *reciprocity principle*, i.e., the unimodularity of the local transfer matrix, the eigenvalues $\Lambda_1$ and $\Lambda_2$ appear as reciprocal pairs associated with the positive and negative directions of the system. Therefore, these can be expressed as,

$$\Lambda_1 = e^{i\mu(\Omega)}, \quad \Lambda_2 = e^{-i\mu(\Omega)} \qquad (23)$$

where, $\mu$ represents the *propagation constant*, which determines the amplitude ratio between motions in adjacent unit cells as a wave traverses the infinite periodic lattice. This constant is frequency-dependent and typically complex. The real part of $\mu$ indicates the phase shift, while the imaginary part reflects the logarithmic decay of amplitude between two successive unit cells. Using the first principal invariant, i.e., the trace of the transfer matrix, $\boldsymbol{T}$, the propagation constant, $\mu$, can be directly expressed as:

$$\mu(\Omega) = arccos\left(\frac{1}{2}tr(\boldsymbol{T})\right) \qquad (24)$$

As expected, by combining equations (23) and (24), the eigenvalues of the *local transfer matrix*, $\boldsymbol{T}$, are fully determined by its trace, which from (18) is expressed as:

$$\begin{aligned}
tr(\boldsymbol{T}) &= T_{11} + T_{22} \\
&= 2cos(k_0(\Delta_1 + \Delta_2)) + sin(k_0(\Delta_1 + \Delta_2))\left(\frac{\sigma_A}{\sigma_a} + \frac{\sigma_B}{\sigma_a}\right) \\
&+ sin(k_0\Delta_1)sin(k_0\Delta_2)\frac{\sigma_A \sigma_B}{\sigma_a^2}
\end{aligned} \qquad (25)$$

For a system without dissipation, where $\lambda_A$ and $\lambda_B$ are both zero, wave propagation is possible if $\mu$ is strictly real (see equation (23)). This requirement is ensured when $-2 \leq tr(\boldsymbol{T}) \leq 2$. The eigenvalues of the transfer matrix that allow for wave propagation, meaning *oscillatory modes*, satisfy the condition $|\Lambda_{1,2}(\Omega)| = 1$. The corresponding frequency bands are bounded by the frequencies at which $tr(\boldsymbol{T})|_{\Omega=\Omega_b} = \pm 2$, where $\Omega_b$ denotes the bounding frequenciy. In contrast, when $\mu = i\theta$ or $\mu = \pi + i\theta$, where $\theta$ is a non-zero real number, band gaps occur. These gaps are present if $tr(\boldsymbol{T}) < -2$ or $tr(\boldsymbol{T}) > 2$. In these cases, one eigenvalue will satisfy $|\Lambda_{1,2}(\Omega)| > 1$, while the other $|\Lambda_{1,2}(\Omega)| < 1$, due to the unimodularity of $\boldsymbol{T}$. For a dissipative system ($\lambda_A, \lambda_B \neq 0$), $\mu$ has both non-zero real and imaginary parts, indicating the existence of complex modes. Table I provides a summary roadmap for characterizing the band structure of the infinite periodic structure.

| | Dissipation | Propagation constant | Eigenvalues | Frequency range |
|---|---|---|---|---|
| Propagation zones | $\lambda_A = \lambda_B = 0$ | $Im(\mu) = 0$ $Re(\mu) \neq 0$ | $|\Lambda_{1,2}(\Omega)| = 1$ | $-2 \leq tr(T) \leq 2$ |
| Band gaps | $\lambda_A = \lambda_B = 0$ | $Im(\mu) \neq 0$ $Re(\mu) = 0 \text{ or } \pi$ | $|\Lambda_1(\Omega)| > 1, |\Lambda_2(\Omega)| < 1$ or $|\Lambda_1(\Omega)| < 1, |\Lambda_2(\Omega)| > 1$ | $tr(T) < -2$ or $tr(T) > 2$ |
| Complex modes | $\lambda_A, \lambda_B \neq 0$ | $Im(\mu) \neq 0$ $Re(\mu) \neq 0$ | $|\Lambda_{1,2}(\Omega)| \neq 1$ | $\forall \Omega$ |

**Table I**: A roadmap for characterization of the band structure for the infinite periodic lattice.

The unimodularity of the local transfer matrix $T$ implies identical qualitative behavior for the band structure for positive- and negative-going waves. Therefore, without loss of generality, the band structure will be analyzed only for the case of negative-going waves, i.e., for positive $\mu$. Since the band structure is determined only by the trace of the *transfer matrix*, $tr(T)$, it can be seen from (25) that the band structure remains robust or insensitive when interchanging between cavities (i.e., $\Delta_1 \leftrightarrow \Delta_2$), as well as when interchanging between membranes (i.e., $\sigma_A \leftrightarrow \sigma_B$). Let $\Omega_{b(n)}^{cut-on}$ and $\Omega_{b(n)}^{cut-off}$ denote the cut-on and cut-off frequencies of the $n^{th}$ propagation band, respectively. As mentioned above, these bounding frequencies, satisfying the relationships $\Omega_{b(n)}^{cut-on} < \Omega_{b(n)}^{cut-off} < \Omega_{b(n+1)}^{cut-on} < \Omega_{b(n+1)}^{cut-on} < \cdots$, are obtained by solving the following transcendental characteristic equation,

$$2cos(k_0(\Delta_1 + \Delta_2)) + sin(k_0(\Delta_1 + \Delta_2))\left(\frac{\sigma_A}{\sigma_a} + \frac{\sigma_B}{\sigma_a}\right) + sin(k_0\Delta_1)sin(k_0\Delta_2)\frac{\sigma_A\sigma_B}{\sigma_a^2} = \pm 2 \quad (26)$$

where $k_0$, $\sigma_a$, $\sigma_A$, and $\sigma_B$ are dependent on $\Omega$. Based on the bounding frequencies definition, the width of the $n^{th}$ band gap, denoted by $W_{(n)}^{gap}$, is simply determined by:

$$W_{(n)}^{gap} = \Omega_{b(n+1)}^{cut-on} - \Omega_{b(n)}^{cut-off} \quad (27)$$

Therefore, the analysis of the band structure of the infinite periodic lattice consisting of the complex multilayer vibroacoustic membrane-cavity resonators, is simplified to the determination of the roots of the transcendental characteristic equation (26). This equation provides the bounding frequencies of the bands, enabling a comprehensive understanding of the physical mechanisms governing the vibroacoustic interactions within the band gaps.

Referring to the characteristic equation (26), both cut-on and cut-off bounding frequencies for the $n^{th}$ band/gap generally demonstrate complex nonlinear dependencies on the cavities' depths ($\Delta_1$ and $\Delta_2$), air properties ($\sigma_a$), membranes' properties ($\sigma_A$ and $\sigma_B$), and the unit-cell length ($\Delta_1 + \Delta_2$). Therefore, prior to analyzing the band structure of a general unit-cell (comprising different membranes and cavities), we will first consider simplified versions of the unit-cell. Lastly, the finite periodic lattice can be analyzed through the construction of the global transfer matrix. We refer to [52] for a general explicit formalism of the global transfer matrix in terms of the local transfer matrix, based on the Cayley–Hamilton theorem.

## III. RESULTS AND DISCUSSION

In the context of the previous multilayered vibroacoustic analysis, four different unit-cell configurations can be considered, namely, *Monolayered* (identical cavities and membranes within a unit cell), *Multilayered type I* (identical cavities but different membranes), *Multilayered type II* (different cavities but identical membranes), and *Multilayered type III* (different cavities and membranes). A 2D schematic representation of infinite periodic lattices corresponding to these configurations is illustrated in Figure 2. The Monolayered vibroacoustic unit-cell was recently analyzed, and its capacity and limitations for low-frequency sound manipulation were outlined [52]. Therefore, this current paper is entirely devoted to providing a complete characterization of the Multilayered cases, aiming to achieve novel capabilities beyond those of the Monolayered case, specifically in the ultra-low frequency domain.

Unless stated otherwise, throughout this study the system parameters listed in Table II are considered. We used the actual values for air speed ($c_a$) and density ($\rho_a$) and selected Polydimethylsiloxane (PDMS) as the membrane material. The speed of elastic wave propagation in the membranes ($c_A$ and $c_B$) and the depths of the cavities ($\Delta_1$ and $\Delta_2$) are treated as design parameters to investigate the band structure evolution behavior. In practice, controlling the wave velocities ($c_A$ and $c_B$) of the membranes, given their material and geometry, is achieved by adjusting their (uniform) tension.

| Parameter | Description |
|---|---|
| $c_a = 343 \ [m/s]$ | Speed of sound in air |
| $\rho_a = 1.255 \ [kg/m^3]$ | Air density |
| $h_A = h_B = 0.0011 \ [m]$ | Thicknesses of membranes A and B |
| $\rho_A = \rho_B = 1050 \ [kg/m^3]$ | Densities of membranes A and B |
| $\lambda_A = \lambda_B = \begin{cases} 0 & lossless \\ 30 & lossy \end{cases} \ [sN/m^3]$ | Viscous damping coefficients of membranes A and B |
| $a = 0.05 \ [m]$ | Membranes' radius |

**Table II**: System parameters that were kept fixed throughout the current study.

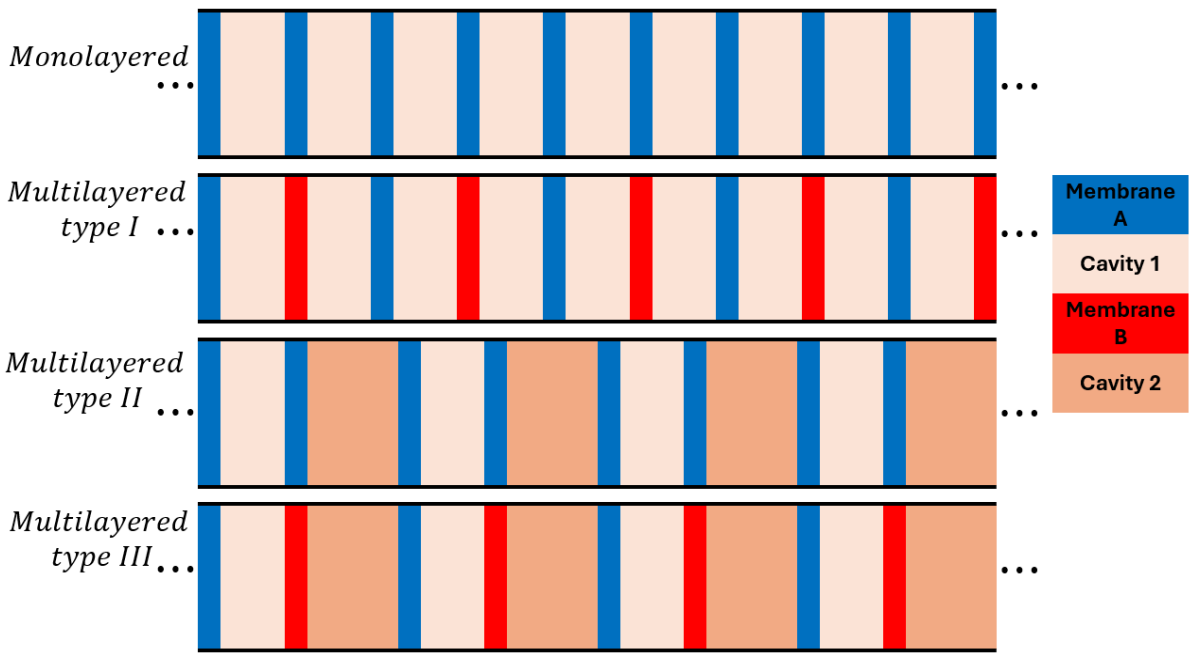

**Figure 2:** Schematic representation of four different configurations of the proposed multilayered vibroacoustic metamaterial: Monolayered ($\Delta_1 = \Delta_2$, $\sigma_A = \sigma_B$), Multilayered type I ($\Delta_1 = \Delta_2$, $\sigma_A \neq \sigma_B$), Multilayered type II ($\Delta_1 \neq \Delta_2$, $\sigma_A = \sigma_B$), and Multilayered type III ($\Delta_1 \neq \Delta_2$, $\sigma_A \neq \sigma_B$).

Regardless of the membranes, which can be regarded as internal resonators within each unit-cell, Bragg diffraction occurs when an integer number of half-wavelengths fits into a single unit-cell. Focusing on the low-frequency domain, the first-order simple Bragg diffraction frequency for the different Multilayered unit-cell configurations is given by:

$$\Omega_{Bragg}^{(1)} = \frac{1}{2\pi} \frac{2\pi c_a}{(\Delta_1 + \Delta_2)} \ [Hz] \tag{28}$$

At the frequency $\Omega_{Bragg}^{(1)}$, Bragg bandgaps are expected to form. Additionally, as local resonators, the membranes contribute to the formation of local resonance bandgaps. Unlike Bragg bandgaps, however, these local resonance bandgaps are expected to form around the membranes' natural frequencies, regardless of the structure's (global) periodicity. According to equation (14), the $n^{th}$ undamped and damped natural frequencies of Membrane A(B) interacting with the cavities, respectively denoted by $\Omega_{A(B)}^{(n)}$ and $\Omega_{A(B),lossy}^{(n)}$, can be explicitly expressed as:

$$\Omega_{A(B)}^{(n)} = \frac{c_{A(B)}}{a} \gamma_2^{(n)}, \quad \Omega_{A(B),lossy}^{(n)} = \frac{i\lambda_{A(B)}}{2\rho_{A(B)} h_{A(B)}} + \sqrt{\left(\frac{c_{A(B)}}{a}\gamma_2^{(n)}\right)^2 - \left(\frac{\lambda_{A(B)}}{2\rho_{A(B)} h_{A(B)}}\right)^2} \tag{29}$$

Here, $\gamma_2^{(n)}$ is the $n^{th}$ root of the second-order Bessel function of the first kind. These relations yield a countable infinity of natural frequencies for the in-air membranes.

### A. Overview of sub-Bragg phenomena in multilayered vibroacoustic metamaterials

Representative band structures for each multilayered configuration as well as the monolayered case (see Figure 2), are illustrated in Figure 3, encompassing both undamped and damped cases. Specifically, the real and imaginary parts of the propagation constant, $\mu$, for the four leading pass bands (represented by cyan strips) and bandgaps (marked with gray-scale shading) are presented. Across the four multilayered configurations, the unit cell length ($\Delta_1 + \Delta_2$) is fixed at 0.1 $[m]$. According to equation (28), the first-order Bragg frequency for these configurations is $\Omega_{Bragg}^{(1)} = 3430\ [Hz]$. Since the frequencies of interest in Figure 3 are below 600 $[Hz]$, the bandgaps observed cannot be attributed to classical Bragg diffraction and are therefore considered *sub-wavelength bandgaps*. Beyond classical Bragg diffraction and based on the attenuation behavior described by $Imag(\mu)$, the observed bandgaps can be classified into three distinct categories, as indicated by varying levels of gray-scale shading. At this stage, we will briefly introduce these categories, with detailed characterization and physical explanations provided subsequently in this section. The light-gray shading represents *plasma bandgaps*, which consistently appear at the beginning of the band structure regardless of the multilayer configuration. This is a characteristic signature of the membrane-cavity interaction. The intermediate-gray shading denotes *local resonance bandgaps*, which consistently form around the natural frequencies of the membranes (blue lines for Membrane A and red lines for Membrane B) acting as local resonators. These bandgaps are typically characterized by narrow yet strong attenuation performance.

The third category, marked in black, interestingly exhibits Bragg-like attenuation behavior at sub-Bragg frequencies, contrasting with the sub-wavelength local resonance bandgaps, which show a distinct attenuation profile unrelated to Bragg behavior. While these might initially appear to be band-folding induced bandgaps, we will later reveal that band-folding induced bandgaps are a subset of a broader category of novel *sub-Bragg bandgaps*, which we term **band-splitting induced bandgaps**. To our knowledge, this novel class of *sub-Bragg band-splitting induced bandgaps* has not been previously reported. The physical mechanism behind these bandgaps cannot be fully explained by a single band structure and will be elucidated further by studying the evolution of the bandgaps across different multilayered configurations later in this section. A more detailed description of *plasma* and *local resonance bandgaps* will also be provided.

The monolayered configuration is achieved when the original unit-cell, defined in Figure 1, consists of identical membranes and identical cavities, effectively representing two monolayered unit-cells. Consequently, in panel (a) of Figure 3, the band-folding points, which originate from the symmetry-enforced degeneracy, are observable. Apart from the *plasma bandgap*, at the low-frequency domain, the sub-wavelength bandgaps in panel (a) are characterized solely by local resonances with high attenuation magnitudes (i.e., high $Imag(\mu)$) but narrow widths. By breaking this degeneracy—either by using different membranes (i.e., multilayered type I, as shown in panel (b)) or different cavities (i.e., multilayered type II, as shown in panel (c))—we can observe the emergence of *band-splitting induced bandgaps*. We also observe a new widening effect in local resonance bandgaps caused by the simultaneous presence of two local resonances within the same bandgap. This scenario (illustrated by the fourth bandgap in panels (b) and (d)) provides an optimal case for achieving wide and strongly attenuated bandgaps. To quantify the widening effect of hosting two local resonances within the same bandgap, we compare the third bandgap in panel (a), which forms around the second natural frequency of the membrane (approximately $400\ Hz$) with a width of $25\ Hz$. In panel (b), this becomes the fourth bandgap, with a width of $46\ Hz$, due to the simultaneous presence of two membrane local resonances within the same bandgap. This results in a 1.84-fold increase in the bandgap width. Remarkably, this enhanced widening mechanism is achieved within the sub-wavelength range solely by tuning the membranes' natural frequencies, without relying on the classical Bragg mechanism.

The interaction between local resonance bandgaps and band-splitting induced bandgaps, as shown in panel (d) of Figure 3, results in a super-wide attenuation range that includes an ultra-narrow pass band, creating an acoustical transparency window, as will be demonstrated later in this section. Notably, the attenuation performance of band-splitting induced bandgaps (the black strips) is highly dependent on their width; as the width increases, the attenuation magnitude also becomes stronger. Among the multilayered configurations, the multilayered type III configuration exhibits superior performance in manipulating sound waves in the sub-wavelength domain by providing wide bandgaps characterized by strong attenuation and ultra-narrow pass bands.

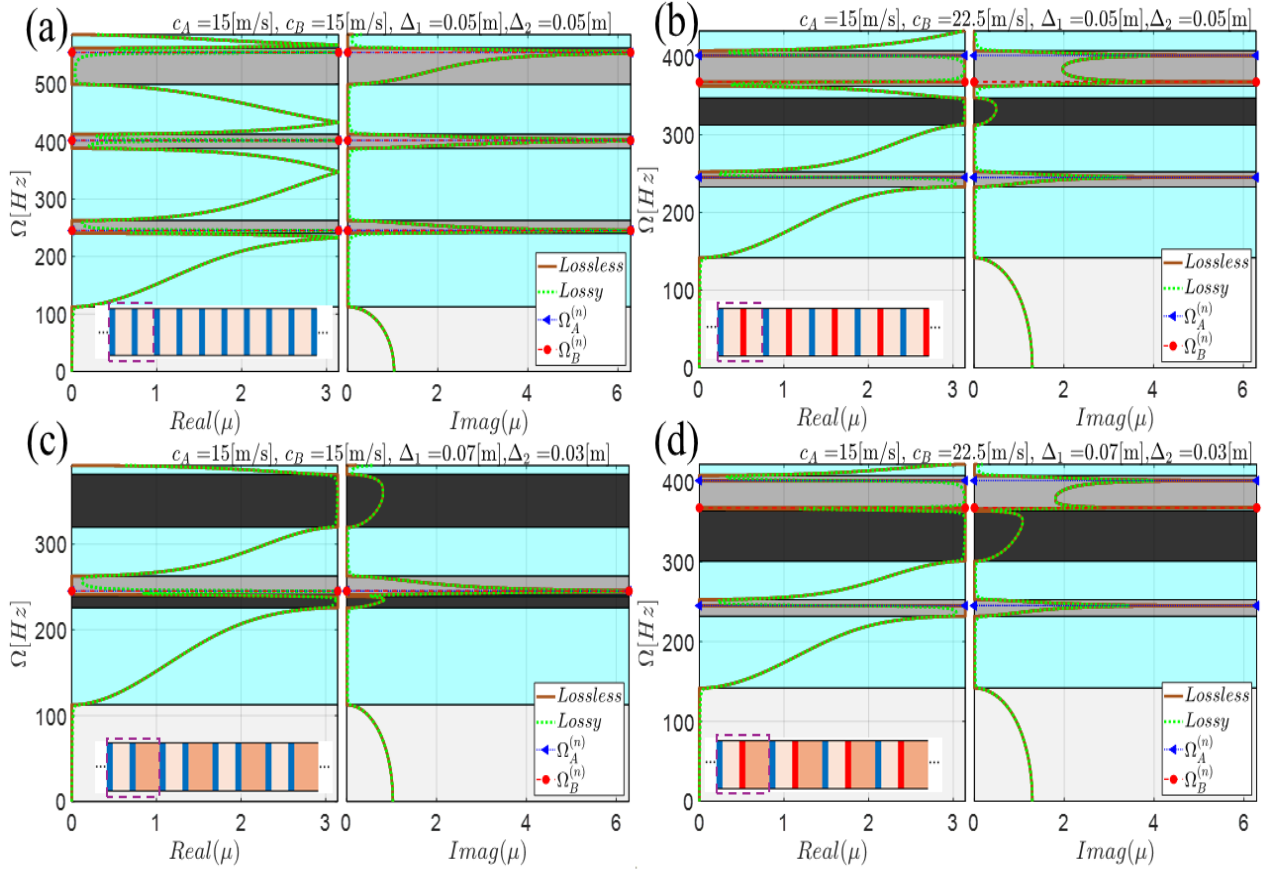

**Figure 3**: The four leading pass bands and bandgaps of the band structure of a typical vibroacoustic metamaterial with infinite number of unit-cells: (top-left) Monolayered with $\Delta_1 = \Delta_2 = 0.05 \ [m]$, $c_A = c_B = 15 \ [m/s]$, (top-right) Multilayered type I with $\Delta_1 = \Delta_2 = 0.05 \ [m]$, $c_A = 15 \ [m/s]$, $c_B = 22.5 \ [m/s]$, (bottom-left) Multilayered type II with $\Delta_1 = 0.07 \ [m]$, $\Delta_2 = 0.03 \ [m]$, $c_A = c_B = 15 \ [m/s]$, and (bottom-right) Multilayered type III with $\Delta_1 = 0.07 \ [m]$, $\Delta_2 = 0.03 \ [m]$, $c_A = 15 \ [m/s]$, $c_B = 22.5 \ [m/s]$. Real and imaginary parts of the propagation constant, $\mu$, given in equation (24), are plotted in the left and right parts of each panel, respectively. The cyan strips indicate the pass bands while the varying levels of gray-scale shading, from light to dark, represent different types of band gaps, which will be classified subsequently. The blue and red lines represent the undamped natural frequencies of membranes A and B, respectively.

### B. Detailed Characterization of the Sub-Bragg Phenomena

To thoroughly characterize and provide detailed physical explanations for the sub-wavelength plasma, local resonance, and band-splitting induced bandgaps, we aim to investigate the evolution of the band structure in infinite multilayered (type I, II, and III) vibroacoustic metamaterials. Additionally, we will discuss the widening mechanism of sub-wavelength local resonance bandgaps and the acoustical transparency introduced earlier. To this end, we consider the bounding frequencies of the bands (26), which directly provide the cut-on and cut-off frequencies of the bands (or the bandgaps). To gain deeper and clearer insights into the pure physical mechanisms behind the formation of bandgaps and their interaction scenarios, we will consider the undamped case, i.e., $\lambda_A = \lambda_B = 0$.

Figure 4 illustrates the effect of the speed of elastic wave propagation in Membrane B ($c_B$) on the evolution of the first six bandgaps across the three different multilayered unit-cell configurations. For a fair comparison, the length of the unit-cell ($\Delta_1 + \Delta_2$) is kept constant across all three configurations. Consequently, according to equation (28), the first-order simple Bragg diffraction frequency is uniform across the three different multilayered configurations and given by $\Omega_{Bragg}^{(1)} = 3430\ [Hz]$. Thus, the formation of the classical Bragg bandgaps is not feasible in the low-frequency range considered in Figure 4. Instead, the obtained bandgaps (indicated by grayscale) are classified as *sub-wavelength bandgaps*, which arise through three different physical mechanisms. The first mechanism is an acoustic analog of *plasma oscillation*, observed in the first bandgap across the three considered configurations, where wave propagation is completely prohibited within the frequency range $0 < \Omega < \Omega_P$. Here, $\Omega_P$ denotes the plasma frequency, determined by the first root of the bounding frequency equation (26). This bandgap, associated with plasma-like behavior, is termed the *plasma bandgap*. Such a phenomenon was experimentally observed in a monolayered vibroacoustic metamaterial [46] and recently analytically investigated by the authors in [52]. It is attributed to the negative effective density behavior exhibited by the membranes within the unit-cell, resulting in high reflectivity below its characteristic plasma frequency $\Omega_P$. *The plasma-like behavior makes the proposed vibroacoustic metamaterial highly effective in isolating ultra-low frequency longitudinal vibrations.* Interestingly, for the Multilayered type II configuration ($c_A = c_B$, $\Delta_1 \neq \Delta_2$) with small cavities such that $\Delta_1, \Delta_2 \ll 2c_a/\Omega$, the plasma frequency can be explicitly approximated as $\Omega_P^{II} \approx \frac{c_A}{2\pi a}\gamma_0^{(1)} [Hz]$, as shown in panel (b) of Figure 4. Here $\Omega_P^{II}$ denotes the Plasma frequency corresponding to the Multilayered type II configuration, and $\gamma_0^{(1)} \approx 2.4048$ is the first root of the zeroth-order Bessel function of the first kind.

The second mechanism for the formation of the sub-wavelength bandgaps shown in Figure 4 is *local resonance bandgaps*, resulting from the membranes acting as local resonators. These local resonance bandgaps emerge around the natural frequencies of membranes A and B, depicted by blue and red lines, respectively, and play a crucial role in shaping the band structure. Typically, local resonance bandgaps in acoustic metamaterials are narrow, each containing only a single local resonance frequency. One way to widen a local resonance bandgap is through its interaction with a Bragg bandgap, resulting in an expanded width near their overlapping range [28-38]. However, such overlap either occurs at high frequencies or requires bulky samples for low frequencies [52]. Another way to achieve a broadband local resonance bandgap is through incorporating multiple arrays resonators in a single unit cell [62, 63]. However, such designs offer only a limited number of bandgaps, and the practical implementation of these artificial structures is highly challenging.

In the multilayered vibroacoustic metamaterial considered herein, *we introduce an additional new mechanism for widening the sub-wavelength local resonance bandgaps, based on the interaction between the natural frequencies of membranes A and B in the unit-cell*. This interaction creates interesting scenarios where two local resonances can exist simultaneously within the same bandgap, significantly widening the sub-wavelength bandgaps, as shown in panels (a) and (c) of Figure 4. The simultaneous existence of the membranes' local resonances within the same bandgap is unique to the multilayered vibroacoustic metamaterial and cannot occur in a monolayered configuration. Additionally, our analytical approach allows for complete and precise control for predictively engineering these intersections between the membranes' local resonances, aiming to enhance the behavior of the metamaterial in the sub-wavelength domain.

Interestingly, our analysis reveals a third mechanism for the formation of sub-wavelength bandgaps, as evidenced by the black strips in the three panels of Figure 4. These strips result from the *splitting of pass bands by the local resonances of the membranes*. As the membrane's local resonance frequency approaches a pass band, the width of the band shrinks progressively until it collapses completely (zero width) due to the intersection with the membranes' local resonances, leading to band splitting. Mathematically, the band-splitting scenario for the $n^{th}$ propagation band collapsing and intersecting with the $m^{th}$ local resonance of Membrane A or B ($\Omega_A^{(m)}$ or $\Omega_B^{(m)}$) is given by $\Omega_{b(n)}^{cut-on} = \Omega_{b(n)}^{cut-off} = \Omega_A^{(m)}$ or $\Omega_B^{(m)}$, where $\Omega_{b(n)}^{cut-on}$ and $\Omega_{b(n)}^{cut-off}$ denote the cut-on and cut-off frequencies of the $n^{th}$ propagation band, respectively. *To our knowledge, this band-splitting phenomenon and its role in generating genuinely sub-wavelength Bragg-like bandgaps is novel and has not been reported in the literature.* We coin the term **"band-splitting induced bandgaps"** to describe these *sub-Bragg* bandgaps. Unlike local resonance bandgaps, these band-splitting induced bandgaps can support band-crossing phenomena (see Figure 4(a)), making them akin to Bragg-like bandgaps. In contrast, band-crossing is not feasible within local resonance bandgaps.

It is worth noting that Figure 4 illustrates two distinct classes of *band-splitting induced bandgaps*. The first class, shown in Figure 4(a), arises from band-crossing phenomena and is known as "band-folding induced bandgaps" [55-58]. However, this mechanism cannot explain the second class of band-splitting induced bandgaps depicted in Figures 4(b) and 4(c), where no band-crossing occurs. Therefore, our new terminology, "*band-splitting induced bandgaps*", provides a new generalization that consistently encompasses both classes of band-splitting phenomena, with "band folding" being a special case within this broader family (see panels (a-c) of Figure 4).

The band-splitting phenomenon, which results in the emergence of Bragg-like band-splitting induced bandgaps in the sub-wavelength region, is achieved by using different membranes, different cavities, or a combination of both within a unit-cell, as shown in panels (a), (b), and (c) of Figure 4, respectively. This phenomenon is exclusive to multilayered configurations and is not feasible in monolayered ones. The advantage of multilayered type III (simultaneously using different membranes and different cavities) over types I and II in obtaining significantly wide band-splitting induced bandgaps in the sub-wavelength region is evident, *rendering the multilayered type III an optimal configuration for effectively manipulating sound waves in the low-frequency domain*. Notably, *the multilayered type III design can entirely annihilate the first pass band*, a result unattainable with the monolayered configuration [52].

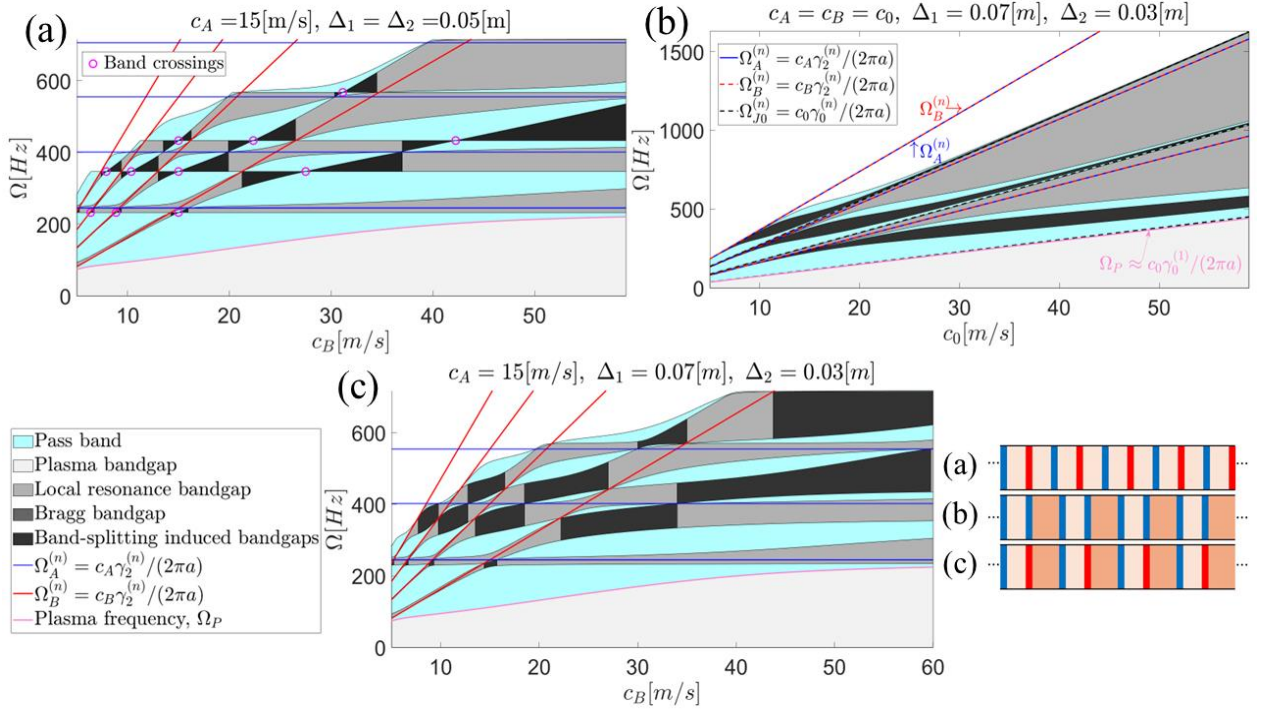

**Figure 4**: Evolution of the six leading bandgaps in the infinite undamped multilayered vibroacoustic metamaterial as the speed of elastic wave propagation in Membrane B ($c_B$) varies: (a) Multilayered type I with $\Delta_1 = \Delta_2 = 0.05\ [m]$, $c_A = 15\ [m/s]$, (b) Multilayered type II with $\Delta_1 = 0.07\ [m]$, $\Delta_2 = 0.03\ [m]$, $c_A = c_B \equiv c_0$, and (c) Multilayered type III with $\Delta_1 = 0.07\ [m]$, $\Delta_2 = 0.03\ [m]$, $c_A = 15\ [m/s]$. The cyan strips indicate the pass bands, while the four levels of gray-scale shading, from light to dark, correspond to the plasma, local resonance, Bragg, and band-splitting induced band gaps, respectively. The blue and red lines represent the undamped natural frequencies of membranes A and B. The solid pink contour illustrates the plasma frequency. Black dashed lines denote the frequencies corresponding to the roots of the zeroth-order Bessel function of the first kind (i.e., the in-vacuo membrane natural frequencies, see equation (12)). Magenta hollow circles highlight the band-crossing points.

The band-crossing points depicted in Figure 4(a) can be classified into two qualitatively distinct categories. The first category comprises symmetry-enforced band crossings, associated with the family of band crossings located along the vertical line where $c_B = c_A$, i.e., corresponding to the monolayered case. The second category consists of band crossings that are unrelated to any symmetry of the system and can thus be classified as accidental band crossings. Another peculiar yet interesting phenomenon occurring in the multilayered vibroacoustic metamaterial is the existence of ultra-narrow pass bands resulting from band-splitting scenarios, which can be produced in the sub-wavelength domain. These ultra-narrow pass bands become densely spaced with resonant transmission when considering a finite version of the periodic lattice. These closely spaced resonant transmission points create *transparency windows* with perfect transmission, which can be considered an acoustic analogue of the well-known electromagnetically induced transparency caused by quantum interference [33]. It is worth noting that the *acoustical transparency* studied in [33] resulted from the interplay between local resonance and classical Bragg bandgaps. In contrast, the *acoustical transparency* observed here occurs at frequencies well below the first-order Bragg diffraction. This *sub-wavelength acoustical transparency* is valuable for achieving precise low-frequency acoustic wave filters and could be applied in advanced noise control, sound manipulation, and signal processing technologies.

The influence of Cavity 2 depth, $\Delta_2$, on the six leading bandgaps for the three multilayered configurations is illustrated in Figure 5. The location of the first-order Bragg diffraction is highlighted in green to emphasize that all bandgaps (plasma, local resonance, and band-splitting induced bandgaps) obtained below the first-order Bragg are sub-Bragg bandgaps. Figure 5 clearly demonstrates that achieving classical Bragg diffraction, even the first-order Bragg, requires relatively bulky unit-cells. Consequently, the current study focuses on achieving sub-wavelength and wide bandgaps without relying on Bragg diffraction. It is evident from Figure 5 that the proposed multilayered configurations, particularly type III, are capable of producing such wide and sub-wavelength bandgaps independently of Bragg diffraction.

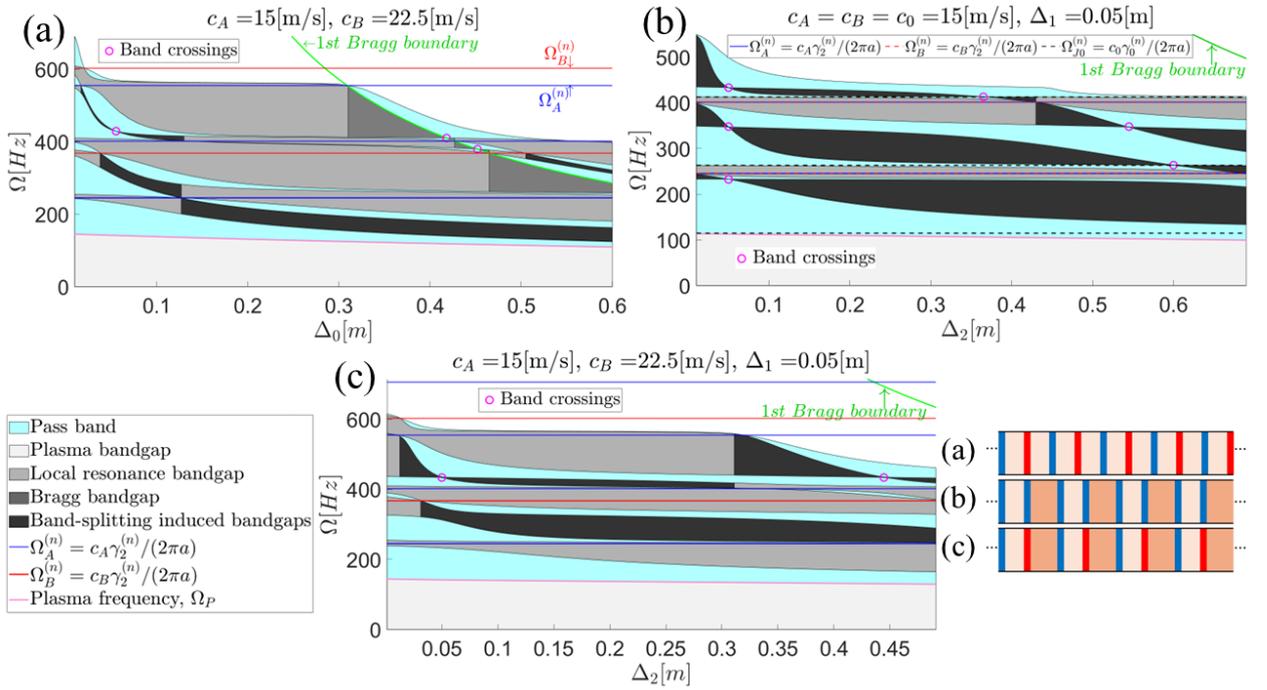

**Figure 5**: Evolution of the six leading bandgaps in the infinite undamped multilayered vibroacoustic metamaterial as the depth of Cavity 2 ($\Delta_2$) varies: (a) Multilayered type I with $\Delta_1 = \Delta_2 \equiv \Delta_0$, $c_A = 15\ [m/s]$, $c_B = 22.5\ [m/s]$, (b) Multilayered type II with $\Delta_1 = 0.05\ [m], c_A = c_B \equiv c_0 = 15\ [m/s]$, and (c) Multilayered type III with $\Delta_1 = 0.05\ [m]$, $c_A = 15\ [m/s]$, $c_B = 22.5\ [m/s]$. Solid green curve represents the first-order Bragg frequency boundary. The interpretations of the colors are the same as described in Figure 4.

To investigate the role of asymmetry in the depth of the two cavities, we define a detuning parameter $\gamma$ such that the depth of each cavity is given by:

$$\begin{aligned}\Delta_1 &= \Delta_0(1+\gamma),\\ \Delta_2 &= \Delta_0(1-\gamma)\end{aligned} \quad (30)$$

where $\Delta_0$ is a nominal cavity depth, and $\gamma$ represents the deviation from symmetry in the cavity depths. Notably, by solely changing the detuning parameter, $\gamma$, the length of the unit-cell remains constant and equal to $2\Delta_0$. Figure 6 illustrates the evolution of the six leading bandgaps for multilayered types II and III for varying detuning parameter, $\gamma$, while keeping the unit-cell length and the membranes' wave speeds, $c_A$ and $c_B$, unchanged. It is evident that the symmetric case ($\gamma = 0$) results in narrower bandgaps compared to the strongly asymmetric arrangement when $\gamma \to 1$. This highlights the significant role of asymmetric cavities in achieving wide and sub-wavelength bandgaps in the low-frequency domain.

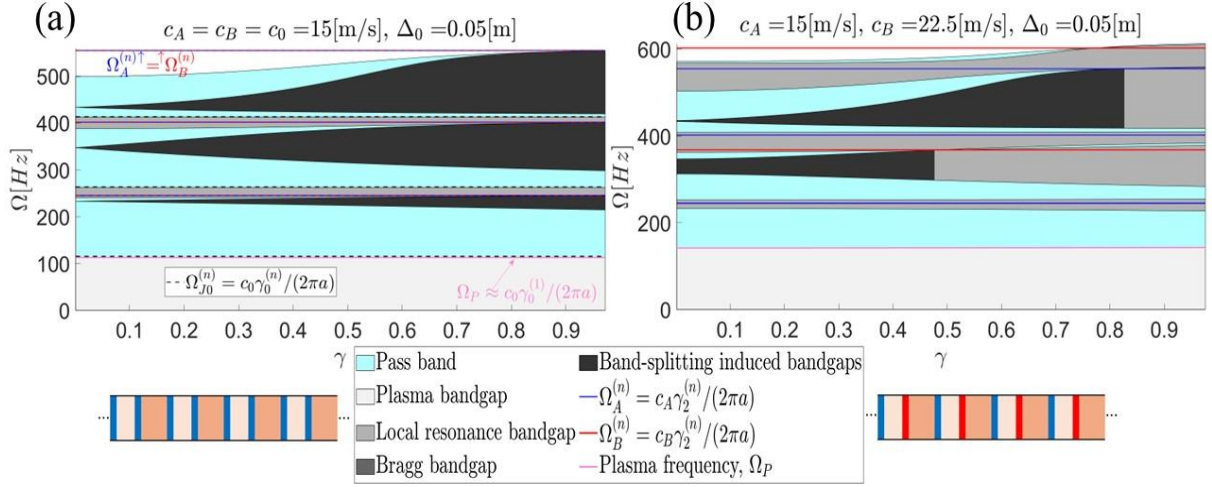

**Figure 6**: Evolution of the six leading bandgaps in the infinite undamped multilayered vibroacoustic metamaterial as the contrast parameter $\gamma$ (see equation (30)) varies: (a) Multilayered type II with $\Delta_0 = 0.05$ $[m]$, $c_A = c_B \equiv c_0 = 15$ $[m/s]$, and (b) Multilayered type III with $\Delta_0 = 0.05$ $[m]$, $c_A = 15$ $[m/s]$, $c_B = 22.5$ $[m/s]$. The interpretations of the colors are the same as described in Figure 4.

### IV. CONCLUDING REMARKS

To achieve enhanced sound wave manipulation in the low-frequency range, we proposed a vibroacoustic phononic metamaterial consisting of periodic multilayered unit cells, each comprising two layers of membrane-cavity resonators. Initially, under the assumption of axisymmetric modes, we developed an analytical solution for the coupled vibroacoustic system for a representative unit cell. Our approach yielded exact closed-form solutions for the transverse displacement fields of the membranes and the longitudinal pressure fields within the cavities, fully accounting for the sound-membranes interactions effects. Next, by employing the Bloch-Floquet theorem in conjunction with the transfer matrix method, we performed a comprehensive analytical characterization of the infinite phononic system's band structure, including the bounding frequencies of bandgaps and dispersion branches. Notably, we derived a single transcendental characteristic equation for determining the bounding frequencies of the passbands and bandgaps of the metamaterial. Subsequently, we investigated the influence of the wave speeds of the membranes, the cavity depths, and the asymmetry of the unit cell on the evolution of the band structure across three different multilayered configurations.

Our analysis revealed new sub-Bragg mechanisms beyond the classical Bragg diffraction and local resonance bandgaps, specifically the widening of sub-wavelength local resonance bandgaps, and forming genuinely sub-wavelength Bragg-like band-splitting induced bandgaps. The wide sub-wavelength local resonance bandgaps are achieved by hosting two local resonances within the same bandgap, resulting in superior wide bandgaps with strong attenuation. The band-splitting phenomenon occurs when a membrane local resonance intersects with a pass band, resulting in the splitting of the band and the creation of bandgaps characterized by Bragg-like attenuation behavior in the sub-wavelength range, which were termed band-splitting induced bandgaps. These bandgaps represent a generalization of the known band-folding induced bandgaps by introducing the possibility of either containing band-crossing points or eliminating them entirely, thereby expanding the range of possible bandgap structures. Additionally, the band-splitting phenomenon creates ultra-narrow pass bands, leading to sub-wavelength transparency windows with perfect transmission, akin to the well-known electromagnetically induced transparency, unlike traditional acoustical transparency that relies on Bragg diffraction. Furthermore, the current vibroacoustic metamaterial exhibits an acoustic analogue of plasma oscillations, observed in the first bandgap, where wave propagation is entirely prohibited below the plasma frequency, resulting in a plasma bandgap.

Our model also reveals a unique class of accidental band-crossings, unrelated to system symmetry, indicating the potential for non-trivial topological properties. Given our thorough analytical characterization, exploring the extension of this work to topological vibroacoustic modes presents an intriguing direction for future research. Overall, these combined sub-Bragg characteristics make our multilayered vibroacoustic metamaterial highly effective in manipulating low-frequency sound waves. Our findings extend beyond sound propagation, applying to a broader range of physical systems where sub-wavelength behavior is significant.

## ACKNOWLEDGEMENTS

The authors are grateful to the Israeli Council of Higher Education (CHE-VATAT), and the Department of Mechanical Science and Engineering (MechSE) at the University of Illinois Urbana-Champaign for financial support.

## DATA AVAILABILITY STATEMENT

This manuscript has no associated data.

## DECLARATIONS CONFLICT OF INTEREST

The authors declare that they have no conflict of interest.